\documentclass[aps,twocolumn,nofootinbib,superscriptaddress]{revtex4-1}
\usepackage{amssymb}
\usepackage{amsmath}
\usepackage{amstext}
\usepackage{amsfonts}
\usepackage{bbold}
\usepackage{slashed}
\usepackage{color}

\def\l{\left}
\def\r{\right}
\def\fr{\frac}
\def\la{\label}
\def\d{\partial}
\def\dd{{\rm d}}

\def\be{\begin{eqnarray}}
\def\ee{\end{eqnarray}}
\newcommand{\p}{\bar{P}}  
  
\newcommand{\D}{{\mathcal D}} 

\begin{document}

\title{Wess-Zumino Terms for Relativistic Fluids, Superfluids, Solids, and Supersolids}

\author{Luca V. Delacr\'{e}taz}
\affiliation{Physics Department and Institute for Strings, Cosmology, and Astroparticle Physics,\\
  Columbia University, New York, NY 10027, USA}
\affiliation{Institut de Th\'eorie des Ph\'enom\`enes Physiques, \\EPFL,
Lausanne, Switzerland}
\author{Alberto Nicolis}
\affiliation{Physics Department and Institute for Strings, Cosmology, and Astroparticle Physics,\\
  Columbia University, New York, NY 10027, USA}
\author{Riccardo Penco}
\affiliation{Physics Department and Institute for Strings, Cosmology, and Astroparticle Physics,\\
  Columbia University, New York, NY 10027, USA}
\author{Rachel A. Rosen}
\affiliation{Physics Department and Institute for Strings, Cosmology, and Astroparticle Physics,\\
  Columbia University, New York, NY 10027, USA}

\begin{abstract}
We use the coset construction of low-energy effective actions to systematically derive Wess-Zumino (WZ) terms for fluid and isotropic solid systems in two, three and four spacetime dimensions.  We recover the known WZ term for fluids in two dimensions as well as the very recently found WZ term for fluids in three dimensions.  We find two new WZ terms for supersolids that have not previously appeared in the literature.  In addition, by relaxing certain assumptions about the symmetry group of fluids we find a number of new WZ terms for fluids with and without charge, in all dimensions.  We find no WZ terms for solids and superfluids. 
\end{abstract}

\maketitle

\noindent
{\bf Introduction.---}
Recently, there has been much interest in understanding the effects of quantum anomalies in hydrodynamics.  In particular, chiral anomalies in a microscopic theory are claimed to lead to new model-independent transport phenomena, once that theory is heated up and brought to a fluid state with finite chemical potentials \cite{Son:2009tf}. The recent interest is due both to the universality of such phenomena---the new transport coefficients are uniquely determined by the anomaly coefficient and by thermodynamic variables---and, perhaps more pragmatically, to the realization that such effects could be observable in heavy-ion collisions \cite{Kharzeev:2010gr}. 

Even though they appear at the same order in the derivative expansion as dissipative effects, the effects in question are non-dissipative.  If one artificially sets to zero the viscosity and heat conduction coefficients, but not the anomaly coefficient, one is left with entropy-conserving ``anomalous transport.''  It is thus natural to ask whether such effects can be captured by a low-energy effective field theory (EFT) description of hydrodynamics~\cite{Dubovsky:2011sj}, which is non-dissipative by construction. In this framework, anomalous effects are described by terms in the action that are invariant only up to a total derivative. These are known as \emph{Wess-Zumino (WZ) terms}~\cite{Wess:1971yu,Witten:1983tw}, and they also belong in the low-energy invariant action~\cite{DHoker:1994ti}. The systematic derivation of such terms is the main focus of this Letter. 

To derive WZ terms, we adopt a standard coset construction.  This formalism is general enough to accommodate not only ordinary fluids, but also superfluids, solids, and supersolids.   We carry out a search for WZ terms for all of these systems in two, three and four spacetime dimensions.   With these methods, we are able to construct new WZ terms.  In particular, for fluids and supersolids we find WZ terms that, to our knowledge, have not appeared previously in the literature.  

The physical effects of these terms will be worked out in future work. In particular, as a first step we will need to translate our WZ terms---which appear as corrections to the {\em action}---into the standard language of ``constitutive relations", that is, as corrections to the currents and stress-energy tensor of the system (see e.g.~\cite{Son:2009tf, Valle:2012em, Jensen:2012kj}). The translation will entail using the Noether theorem and adding improvement terms to restore gauge-invariance, as done for instance in \cite{PhysRevD.89.045016} for (1+1)D fluids.

This Letter is a continuation of the work done in \cite{PhysRevD.89.045002}.  For further details, derivations and conventions we refer the reader to this previous paper. For a different approach to anomalies in hydrodynamics, see instead~\cite{Nair:2011mk}.\\
 \vspace{-.3cm}

 \noindent
{\bf The coset construction.---} From an EFT point of view, relativistic fluids, solids, superfluids and supersolids are not that different from each other.  In all these systems, boosts are spontaneously broken together with other spacetime and internal symmetries. The only symmetries that remain unbroken are large-scale isotropy\footnote{For simplicity we neglect crystalline solids, for which isotropy is replaced by invariance under a discrete subgroup of rotations. Our techniques can be extended straightforwardly to cover this case as well.}
and homogeneity which, depending on the system, can sometimes be associated with  linear combinations of Poincar\'e and internal generators. A concise summary of the symmetry properties of fluids, solids, superfluids and supersolids is provided in Table \ref{summary}.  

At sufficiently low-energies, the effective actions for these systems contain only the Goldstone excitations associated with the relevant symmetry breaking pattern,
%
and can be obtained systematically using the coset construction~\cite{Callan:1969sn,Coleman:1969sm,Volkov:1973vd,Ogievetsky:1974ab}.  Here we present a concise review of the coset construction ``algorithm.''  The starting point is the coset parametrization
\be
\Omega = e^{i x^\mu \bar{P} _\mu} e^{i \pi^a X_a},
\ee
where the $X_a$'s are the generators that are broken by the ground state of the system, the $\pi^a$'s are the associated Goldstone fields, and the $\bar P_\mu$'s are the generators of {\em unbroken} translations, which once again may or may not be linear combinations of Poincar\'e and internal generators. Using $\Omega$, one can build the Maurer-Cartan one-form
\be \la{MC}
\Omega^{-1} d\Omega = i (  \omega_{\bar{P}}^\mu \p_\mu+  \omega_X^a X_a + \omega_T^A T_A ),
\ee
where the $T_A$'s are all the unbroken generators that are not translations. The one-forms $\omega_{\bar P}^\mu$ are related to the spacetime vielbeins by $\omega_{\bar{P}}^\mu = e_\alpha{}^{\mu} \, dx^\alpha$, while the one-forms $\omega_X^a $ yield the covariant derivatives of the Goldstone fields via $\omega_X^a = e_\alpha{}^{\nu} \,  \D_\nu \pi^a  \, dx^\alpha$. Finally, the one-forms $\omega_T^A$ can be used to define higher-order covariant derivatives of the Goldstones as well as covariant derivatives of matter fields.  Together, these building blocks can be used to create generic effective Lagrangians that are invariant under the full symmetry group, with the broken symmetries realized non-linearly.  

As is well known, the relativistic Goldstone theorem does not directly apply to relativistic systems that spontaneously break \emph{spacetime} symmetries.  In particular, the number of broken generators can exceed the number of low-energy excitations.  In the coset construction, this mismatch arises because certain covariant derivatives $ \D_\nu \pi^a$ can be set to zero while preserving all the symmetries and such conditions can be solved to eliminate some of the Goldstone fields from the Lagrangian. 
These conditions are known as \emph{inverse Higgs constraints}~\cite{Ivanov:1975zq}. 
As a consequence of such constraints, our fluid and solid systems are all described in $D$ spacetime dimensions by \emph{at most} $D$ massless Goldstones, even though they all have \emph{at least} $D$ broken symmetries~\cite{PhysRevD.89.045002}. \\
 \vspace{-.3cm}

\begin{table}[b]
\vspace{-.4cm}
\centerline{\footnotesize 
\begin{tabular}{ccccc}
System & $P_0$ & $P_i$ & $J_i$ & Internal symmetries \\
\hline
\bf Superfluid & &  \checkmark&  \checkmark & $U(1)$ \\
\hline
\bf Solid &  \checkmark & & & $ISO(d)$ \\
\hline
\bf Fluid &  \checkmark & & & $ISL(d) \to $ {\it Diff}$\,'(d)$\\
\hline
\bf Supersolid & & & &$U(1)\times ISO(d)$  \\
\hline
\bf Fluid w/ $Q_0$ & & & & $\subset ISL(d+1) \to $ {\it Diff}$\,'(d+1)$  \\
\hline
\end{tabular}
}
\caption{\small \it A checkmark indicates that a translation or rotation generator is unbroken. The last column contains the internal symmetry group which gets spontaneously broken by the ground~state. We are denoting with {\it Diff}$\,'(d)$ the group of diffeomorphisms with unit Jacobian in $d$ dimensions.} \la{summary} 
\end{table}

 \noindent
{\bf Wess-Zumino terms.---}
The coset construction described thus far generates all the terms in the effective Lagrangian that are exactly invariant under the non-linearly realized symmetries.  However, it does not capture Wess-Zumino (WZ) terms~\cite{Wess:1971yu,Witten:1983tw}, which are invariant only up to a total derivative. As mentioned in the introduction, these terms are necessary to describe low-energy consequences of anomalies in the UV theory.  

Wess-Zumino terms can be constructed from the one-forms that are produced by the coset construction.  Let us outline the procedure.  (More details can be found in \cite{DHoker:1994ti,Goon:2012dy}.) For a symmetry group $G$ spontaneously broken down to the subgroup $H$, we can construct exact, invariant $(D+1)$-forms on $G/H$ using the covariant $\omega$'s.  Let us denote such a $(D+1)$-form by $\alpha = d\beta$.  The $D$-form $\beta$ itself is not necessarily invariant but, since $\alpha$ is invariant, it can shift by a total derivative.  We can find all $D$-forms that shift by a total derivative by first finding all exact, invariant $(D+1)$-forms $\alpha$ and then identifying those which differ only by the derivative of a $D$-form $\gamma$ which is itself invariant, i.e. $\alpha \sim \alpha +d\gamma$. Any member of this equivalence class can be identified with the $D$-form $\beta$ that shifts as a total derivative.  The Wess-Zumino terms are obtained by integrating all the inequivalent $D$-forms $\beta$ on the $D$-dimensional spacetime manifold.  



In our analysis we discard tadpole WZ terms because we assume that the ground state corresponds to a solution of the EFT field equations. However, these tadpoles might be relevant for cosmological applications, where the background solutions to the field equations are typically time-dependent.

Some of the WZ terms that we find are invariant under Poincar\'e only up to a total derivative.  These terms are likely to introduce gravitational anomalies upon coupling to gravity.  It would be interesting to understand under what conditions these anomalies can be cancelled by another sector which is somewhat decoupled from the system under consideration.  While our construction allows us to consider such issues, these questions will be studied in detail elsewhere.   For this paper we choose to focus our attention on the remaining WZ terms.

Coset construction techniques are potentially unwieldy when applied to fluids as the low-energy effective action for fluids is invariant under an infinite-dimensional internal symmetry group~\cite{Dubovsky:2011sj}.  However, if one assumes invariance only under a particular finite dimensional subgroup of these symmetries, then the remaining symmetries arise accidentally at lowest order in derivatives~\cite{PhysRevD.89.045002}.  In this Letter, we will follow this  approach.  However, since WZ terms in $D>2$ are of higher order in the derivative expansion, it is not guaranteed that they will also be accidentally invariant under the full, infinite-dimensional symmetry group. We check for this invariance explicitly.  In addition, by focusing on a finite dimensional subgroup, it is possible that our analysis overlooks some terms that are \emph{exactly} invariant under the finite-dimensional subgroup we consider, but invariant only up to total derivatives under some of the symmetries we neglected. Being exactly invariant, these terms can still be obtained from the coset construction for the finite dimensional sub-group, but it is not possible to identify them by following the systematic procedure outlined above.

Finally, the fact that the infinite-dimensional symmetry group arises {\it accidentally} at lowest order in derivatives leads us to question whether this group is, in fact, the correct symmetry group for fluids at higher order in derivatives.  Indeed, at lowest order in derivatives one can show that the conservation of the infinitely many associated Noether charges is equivalent to Kelvin's theorem \cite{Dubovsky}, which is violated by higher-derivative effects like viscosity. This further indicates that the infinite-dimensional symmetries in question might not be fundamental symmetries of fluid systems, but only accidental low-energy ones.
For these reasons we also include in our analysis WZ terms that are invariant only under the finite-dimensional subgroup.
\\
 \vspace{-.3cm}


\noindent
{\bf Results.---}
In order to systematically construct WZ terms, we developed the Matlab code \texttt{EZWZ} which implements the procedure described above.  We plan to make the code publicly available in the near future.  In Table \ref{results} we summarize the main results we obtained from this code for fluids, superfluids, solids, and supersolids.  The third, fourth, and fifth columns of the Table display respectively the number of covariant, invariant and invariant exact $(D+1)$-forms that can be constructed for each system. The number of inequivalent WZ terms is displayed in the sixth column.  The next column shows the number of WZ terms that are not tadpoles ($\slashed{\mathrm{t}}$).  The final column gives the number of remaining terms that are exactly Poincar\'e Invariant (PI). In the case of fluids, we are also showing  in parentheses the number of WZ terms that are accidentally invariant under the full infinite-dimensional group of internal symmetries. To the best of our knowledge, most of the WZ terms we found have never been discussed in the literature before. The only exceptions are two WZ terms for the $(1+1)D$ fluid with a conserved charge~\cite{PhysRevD.89.045016,PhysRevD.89.045002}, and the WZ term for $(2+1)D$ fluids, which first appeared in~\cite{Geracie:2014iva} while this work was being finalized.  This term was found to be related to Hall viscosity.  A detailed study of the physical implications of all the other WZ terms is beyond the scope of the present Letter and will appear elsewhere.

Before discussing each WZ term individually, it is useful to define some quantities that will appear repeatedly in what follows. Supersolids and fluids with a conserved charge in $D=d+1$ spacetime dimensions have exactly $D$ Goldstones. It is then convenient to group them into a $D$-vector $\pi^\mu$, but the reader should keep in mind that they do not transform as a multiplet under Lorentz transformations since Lorentz is broken by the medium itself. With this notation, a fluid without conserved charge is such that $\pi^0 =0$. It turns out that the Goldstones always appear in the action through the combinations $\phi^\mu = x^\mu + \pi^\mu$, which transform as scalars under spacetime symmetries. The ``spatial components'' $\phi^i$ can be thought of as comoving coordinates of the medium's volume elements~\cite{Dubovsky:2011sj}, and can be used to define a metric on such internal space
\be
B^{ij} = \d_\mu \phi^i \d^\mu \phi^j.
\ee
One can also define a time-like unit vector
\be
U^\mu \equiv \fr{\varepsilon^{\mu \alpha_1 \ldots \alpha_d}\partial_{\alpha_1}\phi^1\ldots\partial_{\alpha_d}\phi^d}{\sqrt{\det B}},
\ee
that can be interpreted as the $D$-velocity field of volume elements, since by construction it is parallel to the curves of constant $\phi^i$, i.e. $U^\mu \d_\mu \phi^i =0$. Finally, using $U^\mu$ we can define an identically  conserved current with $d$ more derivatives, namely
\be
J^\mu &=& \varepsilon^{\mu\alpha_1 \ldots \alpha_d}\varepsilon_{\nu_0 \nu_1 \ldots \nu_d} U^{\nu_0} \d_{\alpha_1} U^{\nu_1} \ldots \d_{\alpha_d} U^{\nu_d}.
\ee

\begin{table}[t]
\centerline{\footnotesize 
\begin{tabular}{cccccccc}
System & $D$ & cov. & inv. & exact & WZ & $\slashed{\mathrm{t}}$ & $\slashed{\mathrm{t}}+$ PI\\
\hline
			& (1+1) & 4 & 4 & 4 & 2 & 1 & -\\ \bf
Superfluid	& (2+1) & 15 & 5 & 4 & 1 & 0 & - \\ 
			& (3+1)  & 56 & 6 & 5 & 1 & 0 & - \\ \hline
			& (1+1)  &4 & 4 & 4 & 2 & 1 & - \\ \bf
Solid		& (2+1)  &70 & 18 & 9 & 0 & 0 & - \\ 
			& (3+1)  &1287 & 57 & 22 & 0 & 0 & - \\ \hline 
			& (1+1)  & 10 & 10 & 8 & 4 & 2 & 1 \\ \bf
Supersolid 	& (2+1)  & 126 & 36 & 18 & 2 & 1 & 1 \\
			& (3+1)  & 2002 & 98 & 41 & 1 & 0 &-\\ \hline
			& (1+1)  &4 & 4 & 4 & 2  & 1 &-\\ \bf
Fluid 	
			& (2+1)  & 210 &  40 & 18 & 1 & 1 &1 (1) \\
			& (3+1)  & 8568 & 166 & 57 & 1 & 1 &1 (0) \\ \hline
			& (1+1)  &20 & 20 & 13 & 6  & 5 &4 (1) \\ \bf
Fluid w/ $Q_0$ 		& (2+1)  & 715 & 139 & 48 & 2  & 2 &2 (1) \\ 
			& (3+1)  &26334 & 582 & 166 & 1 & 1 &1 (0)  \\  [3pt]
\hline
\end{tabular}
} 
\caption{} \la{results}
\vspace{-.5cm}
\end{table}

\noindent
{\bf $(1+1)D$ supersolid.---}
The coset of a $(1+1)D$ supersolid can be parametrized~as
\be
\Omega = e^{i x^\mu \p_\mu}  e^{i \eta K} e^{i \pi^\mu Q_\mu},
\ee
where $\p_\mu = P_\mu + Q_\mu$, and $K$ and $Q_\mu$ are the generators of boosts and internal shifts respectively. Our algorithm returned a single WZ term, which can be derived from the exact 3-form
\be
\omega_3 = \varepsilon_{\mu\nu} \, \omega_Q^\mu\wedge\omega_Q^\nu \wedge \omega_K.
\ee
%
The corresponding term in the Lagrangian is
\be \la{1+1ss}
\mathcal{L}_{WZ} =  \varepsilon_{\alpha\beta} \phi^{\alpha} \d_\mu \phi^\beta J^\mu \,  .
\ee
This term is invariant under internal shifts only up to a total derivative. Interestingly, this term enters the effective action for supersolids already at lowest order in the derivative expansion, exactly like the WZ term for a $(1+1)D$ fluid with a conserved charge~\cite{PhysRevD.89.045016}. However, the term (\ref{1+1ss}) is not a WZ term for fluids because it is not invariant under the full nonlinear chemical shift symmetry~\cite{Dubovsky:2011sj}. Conversely, the WZ term for fluids is not a WZ term for supersolids because it is exactly invariant under all the symmetries of a supersolid. \\
 \vspace{-.3cm}


\noindent
{\bf $(2+1)D$ supersolid.---}
The coset parametrization for a $(2+1)D$ supersolid is
\be
\label{Om}
\Omega = e^{i x^\mu \p_\mu}  e^{i \eta^i K_i} e^{i \pi^\mu Q_\mu}  e^{i \alpha L} ,
\ee
where $L$ is the generator of internal $SO(2)$ rotations. In this case, a WZ term can be derived from the exact 4-form
\vspace{-.2cm}
\be
\omega_4 = \varepsilon_{ij}  (\omega_{\bar{P}}^0+\omega_Q^0)\wedge\omega_L  \wedge \omega_K^i\wedge\omega_K^j .
\ee
After some manipulations, this 4-form leads to the following term in the Lagrangian for $(2+1)D$ supersolids:
%
%
%
%
\be
\mathcal{L}_{WZ}  = \phi^0 \varepsilon_{ij} \l\{\frac{ \d_\alpha \phi^i\d_\mu\d^\alpha \phi^j }{\sqrt{\det B}} + B^{1/2}_{jk} \d_\mu B^{-1/2}_{ki}   \r\} J^\mu . \quad 
\ee
This term is a 3-derivative correction to the low-energy effective action for $(2+1)D$ supersolids.  It is invariant under $U(1)$ transformations $\phi^0 \to \phi^0 + c$ only up to total derivatives. \\
 \vspace{-.3cm}


\noindent
{\bf $(1+1)D$ fluid with $U(1)$ charge.---}  The coset parametrization for a $(1+1)D$ fluid with a conserved $U(1)$ charge is
\vspace{-.2cm}
\be
\label{Om(1+1)fluid}
\Omega = e^{i x^\mu \p_\mu} e^{i \eta K} e^{i \pi^\mu Q_\mu}  e^{i \theta F} \, ,
\ee
where $F$ is the generator of \emph{linearized} chemical shifts. As already discussed in~\cite{PhysRevD.89.045016,PhysRevD.89.045002}, there is only one WZ term that is accidentally invariant under \emph{full} chemical shifts, and it is a 0-derivative correction to the low-energy effective action. There are also three more WZ terms that are only invariant under linearized chemical shifts. One of them is again a 0-derivative correction and was first derived in~\cite{PhysRevD.89.045002}. The other two are a 0-derivative and a 1-derivative correction coming from the 3-forms
\be
\omega_3 &=& (\omega_{\bar{P}}^0+\omega_Q^0)  \wedge (\omega_{\bar{P}}^1+\omega_Q^1)  \wedge \omega_K \\
\omega_3 &=& (\omega_{\bar{P}}^0+\omega_Q^0) \wedge \omega_K \wedge \omega_F \, ,
\ee
and they read respectively
\be
\mathcal{L}_{WZ}  &=& \varepsilon_{\alpha\beta}\phi^\alpha\d_\mu \phi^\beta J^\mu \\ 
\mathcal{L}_{WZ}  &=& \fr{\d_\lambda \phi^0 \d^\lambda \phi^1}{(\d \phi^1)^2} \varepsilon^{\mu\nu} J_\mu \d_\nu \phi^0 \, .
\ee
%


\noindent
{\bf $(2+1)D$ fluid.---} The coset parametrization for a $(2+1)D$ fluid is
\vspace{-.2cm}
\be
\label{Om(2+1)fluid}
\Omega = e^{i x^\mu \p_\mu} e^{i \eta^i K_i} e^{i \pi^\mu Q_\mu}  e^{i \alpha^{ij} M_{ij}} \, .
\ee
Here, the $M_{ij}$'s are the 3 generators of $SL(2)$ transformations corresponding to internal {\em linear} diffeomorphisms. As mentioned above, only this (finite dimensional) subgroup of area-preserving diffeomorphisms is implemented in the coset construction. Then, only one WZ term can be written, which is generated by the exact 4-form
\be
\omega_4 = \varepsilon_{ij} \varepsilon_{kl} \, (\omega_{\bar{P}}^i+\omega_Q^i)\wedge(\omega_{\bar{P}}^j+\omega_Q^j)\wedge\omega_K^k\wedge\omega_K^l
\ee
Integrating this form gives the contribution to the Lagrangian:
\vspace{-.2cm}
\be
\mathcal{L}_{WZ}  = \varepsilon_{ij}\phi^i\d_\mu \phi^j J^\mu \, .
\ee

Remarkably, this term turns out to be accidentally invariant under full area-preserving diffeomorphisms. Note also that this term is not a WZ term for solids, because it can be made exactly invariant under all the solid symmetries by adding a non-invariant total derivative.

Interestingly, one can also start from a different, equivalent 4-form $\tilde{\omega}_4 \sim \omega_4$ defined as
%
%
%
%
%
\be
\tilde{\omega}_4 = \varepsilon_{ij} \varepsilon_{kl} \delta_{mn} (\omega_{\bar{P}}^i + \omega_Q^i)\wedge(\omega_{\bar{P}}^j+\omega_Q^j)\wedge\omega_M^{(km)}\wedge\omega_M^{(ln)}\, , \nonumber 
\ee
and show that
\be
\tilde{\omega}_4\propto \dd^2\phi\wedge {\rm Tr} \left(\varepsilon \dd G G^{-1}\wedge \dd G \right)\, ,
\ee
where $G=B/\sqrt{\det B}$ and $\dd^2\phi=\frac{1}{2}\varepsilon_{ij}\dd\phi^i\wedge\dd\phi^j$. This term and its connection with Hall viscosity were recently discussed in~\cite{Geracie:2014iva}. 
\\
 \vspace{-.3cm}


\noindent
{\bf $(2+1)D$ fluid with $U(1)$ charge.---}
A fluid that carries a $U(1)$ charge benefits from an additional chemical shift symmetry. Considering only the linearized version of this symmetry, the coset parametrization is
\be
\label{Om(3+1)fluidQ}
\Omega = e^{i x^\mu \p_\mu} e^{i \eta^i  K_i} e^{i \pi^\mu Q_\mu} e^{i \theta^i F_i}  e^{i \alpha^{ij} M_{ij}} \, ,
\ee
where the $F_i$ are the generators of linear chemical shifts.
The term that was found for $(2+1)D$ fluids without charge can also be written here. In addition, another WZ term is generated by the 4-form
\be
\omega_4 = \varepsilon_{ij} \varepsilon_{kl} \, \omega_F^i\wedge\omega_F^j \wedge \omega_K^k\wedge\omega_K^l \, ,
\ee
which gives rise to the following contribution to the Lagrangian
%
\be
\mathcal{L}_{WZ} = \varepsilon_{ij} V^i \d_\mu V^j J^\mu \, ,
\ee
with $V^i = \varepsilon^{\rho\nu\lambda} U_\rho \d_\nu \phi^i \d_\lambda \phi^0$. This is a 3-derivative correction to the hydrodynamics of this system that is only invariant under linearized chemical shift and diffeomorphisms. \\
 \vspace{-.3cm}


\noindent
{\bf $(3+1)D$ fluid.---} For a fluid with or without any conserved charge in $(3+1)D$, the invariant 5-form that gives rise to the single Wess-Zumino term can be written as
\be\la{form 3+1 fluid}
\omega_5  = \omega_M^{(ij)}\wedge\omega_M^{(jk)}\wedge\omega_M^{(kl)}\wedge\omega_M^{(lm)}\wedge\omega_M^{(mi)} = {\rm Tr} \left(G^{-1}\dd G\right)^5 \nonumber 
\ee
%
It is easy to check that this form is only invariant under linearized internal diffeomorphisms, not under full ones. At lowest order in the Goldstone fields, it reads
\be
\mathcal{L}_{WZ} \approx \varepsilon^{\mu\nu\lambda\rho} \d_{(i} \pi_{j)}  \d_\mu  \d_{(j} \pi_{k)}  \d_\nu  \d_{(k} \pi_{l)}  \d_\lambda  \d_{(l} \pi_{m)}  \d_\rho \d_{(m} \pi_{i)} \, , \nonumber 
\ee
where repeated indices are summed over. This is a 4-derivative correction to $(3+1)D$ hydrodynamics. \\
 \vspace{-.3cm}


\noindent
{\bf Outlook.---}  The physical interpretation and consequences of these new WZ terms are interesting open questions, which we will address in a separate paper.  It remains to be understood which (if any) of these terms arise from a quantum anomaly of the underlying microscopic theory, as in the case of the $(1+1)D$ fluid. In particular, it would be interesting to explore the physical origin of anomalies associated with emergent symmetries such as internal diffeomorphisms and chemical shifts, which have no obvious counterparts in the microscopic description of fluids. These symmetries are ``emergent" in that they act on the low-energy collective excitations only, and not on the microphysical degrees of freedom. But, unlike more traditional low-energy accidental symmetries like baryon number, they can still be taken as {\em exact} rather than approximate, at least to all orders in the derivative expansion. This makes their physical interpretation particularly subtle.

For our fluid and solid systems, our analysis exhausts the list of possible WZ terms that can be generated via the standard coset construction. While for spontaneously broken {\em internal} symmetries such a method is known to generate all the possible WZ terms \cite{DHoker:1994ti}, to the best of our knowledge this has not been proven for spontaneously broken spacetime symmetries. But, assuming such a result holds, it is notable that we didn't find a one-derivative WZ term for (3+1)D fluids that reproduces the chiral vortical effect of \cite{Son:2009tf}.  In light of this, it's not yet evident if this term can be derived at zero temperature, as it would appear in \cite{Nair:2011mk}, or if one needs to use a finite-temperature approach such as the Schwinger-Keldysh formalism, as suggested by \cite{Haehl:2013hoa}.\\
 
Finally, it was recently discovered by using a dual gauge theory description that (2+1)D superfluids also admit a WZ term~\cite{Golkar:2014paa}. However, when expressed in terms of the scalar field used in this paper, this term turns out to be invariant up to a total derivative only \emph{on-shell}. Our construction instead is based on the requirement of \emph{off-shell} invariance of the action, which is why we didn't find such a term. A systematic study of these ``on-shell WZ terms'' will be the subject of future work.



\noindent 
{\bf Acknowledgments.---}  We would like to thank S. Dubovsky, S. Endlich, K. Hinterbichler, L. Hui, R. Loganayagam, V.P. Nair, R. Rattazzi and D.T. Son for very stimulating conversations. The work of AN, RP and RAR was supported by NASA under contract NNX10AH14G and by the DOE under contract DE-FG02-11ER41743.

\bibliographystyle{apsrev4-1}
\bibliography{anomalies}

\end{document}